\documentclass[modern]{aastex63}
\defcitealias{McQuillan2014}{MMA14}
\received{}
\revised{}
\accepted{}
\submitjournal{ApJL}

\shorttitle{}
\shortauthors{Reinhold et al.}

\begin{document}

\title{Where have all the solar-like stars gone? Rotation period detectability at various inclinations and metallicities}

\correspondingauthor{Timo Reinhold}
\email{reinhold@mps.mpg.de}

\author[0000-0002-1299-1994]{Timo Reinhold}
\affiliation{Max-Planck-Institut f\"ur Sonnensystemforschung \\ Justus-von-Liebig-Weg 3, 37077 G\"ottingen, Germany}

\author[0000-0002-8842-5403]{Alexander I. Shapiro}
\affiliation{Max-Planck-Institut f\"ur Sonnensystemforschung \\ Justus-von-Liebig-Weg 3, 37077 G\"ottingen, Germany}

\author[0000-0002-0929-1612]{Veronika Witzke}
\affiliation{Max-Planck-Institut f\"ur Sonnensystemforschung \\ Justus-von-Liebig-Weg 3, 37077 G\"ottingen, Germany}

\author[0000-0001-6090-1247]{Nina-E. N{\`e}mec}
\affiliation{Max-Planck-Institut f\"ur Sonnensystemforschung \\ Justus-von-Liebig-Weg 3, 37077 G\"ottingen, Germany}
\affiliation{Institut f\"ur Astrophysik, Georg-August-Universit\"at G\"ottingen, Friedrich-Hund-Platz 1, D-37077 G\"ottingen, Germany}

\author[0000-0001-6163-0653]{Emre I\c{s}{\i}k}
\affiliation{Dept. of Computer Science, Turkish-German University \\
\c{S}ahinkaya Cd. 108, Beykoz, 34820 Istanbul, Turkey}

\author[0000-0002-3418-8449]{Sami K. Solanki}
\affiliation{Max-Planck-Institut f\"ur Sonnensystemforschung \\ Justus-von-Liebig-Weg 3, 37077 G\"ottingen, Germany}
\affiliation{School of Space Research, Kyung Hee University, \\
Yongin, Gyeonggi, 446-701, Korea}




\begin{abstract}
The plethora of photometric data collected by the {\it Kepler} space telescope has promoted the detection of tens of thousands of stellar rotation periods. However, these periods are not found to an equal extent among different spectral types. Interestingly, early G-type stars with near-solar rotation periods are strongly underrepresented among those stars with known rotation periods. In this study we investigate whether the small number of such stars can be explained by difficulties in the period determination from photometric time series. For that purpose, we generate model light curves of early G-type stars with solar rotation periods for different inclination angles, metallicities and (magnitude-dependent) noise levels. We find that the detectability is determined by the predominant type of activity (i.e. spot or faculae domination) on the surface, which defines the degree of irregularity of the light curve, and further depends on the level of photometric noise. These two effects significantly complicate the period detection and explain the lack of solar-like stars with known near-solar rotation periods. We conclude that the rotation periods of the majority of solar-like stars with near-solar rotation periods remain undetected to date. Finally, we promote the use of new techniques to recover more periods of near-solar rotators.
\end{abstract}



\section{Introduction} \label{sec:intro}

Stellar brightness variations at the timescale of stellar rotation are caused by transits of magnetic features (such as dark spots or bright faculae) rotating across the visible disk. These variations have routinely been observed by transit photometry missions. In particular, the {\it Kepler} telescope obtained light curves of roughly 150,000 main-sequence stars. Some of these light curves exhibit clear signatures of stellar rotation, which can be extracted by standard frequency analysis tools such as Lomb-Scargle periodograms, auto-correlation functions, or wavelet transforms. The biggest survey of rotation periods based on the {\it Kepler} data has been published by \citealt{McQuillan2014} (hereafter \citetalias{McQuillan2014}), who detected rotation periods for 34,030 presumably main-sequence stars.

However, for the majority of the main-sequence stars the light curves were either too noisy or too irregular for the rotation period to be determined. \citetalias{McQuillan2014} found that the fraction of stars with detectable periods strongly depends on the effective temperature. Interestingly, this fraction appeared to be lowest for stars with near-solar effective temperature (between 5500--6000\,K, hereafter referred to as early G-type stars), reaching only 16\% (see Table~3 in \citetalias{McQuillan2014}). On the contrary, \citet{vanSaders2019} used Galactic evolution models to predict that $\sim$59\% of the early G-type stars should have detectable rotation periods.  

The lack of stars with known rotation period becomes even more severe for early G-type stars of near-solar age. Recently, \citet{Reinhold2020} showed that only a dozen {\it Kepler} stars with near-solar fundamental parameters and rotation periods between 20--30~days (i.e. encompassing the solar rotation period of $\sim$25~days) exhibit rotational variability levels similar to that of the Sun. In contrast, the majority of these stars are substantially more variable than the Sun and also show more regular light curves patterns. It has been proposed that such a conspicuous difference between the Sun and other solar-like stars can be explained by a detection bias towards more active stars in bulk rotation period measurements (see the discussion in \citealt{Eliana2020}), thus missing the majority of early G-type stars with near-solar rotation periods and small variabilities.

In this Letter we address the question whether the "missing" solar-like stars (i.e. stars with solar fundamental parameters and rotation periods) do not exist or simply go undetected. Our approach is based on the solar paradigm, i.e. we build on the comprehensive understanding of solar brightness variability (see, e.g., reviews by \citealt{TOSCA2013, MPS_AA}), and extend solar models to solar-like stars. Namely, we combine two recently developed physics-based models by \citet{Witzke2020} and \citet{Nemec2020}. This allows calculating light curves of stars with a solar distribution of active regions and solar effective temperature, but various metallicities and observed at arbitrary inclination angles. These light curves are used to identify obstacles in the period determination of solar-like stars, and to discuss possible limitations of period measurements in real data sets. We further compare the number of actual period measurements in {\it Kepler} data to predictions from Galactic evolution models using the detection rate obtained from the model light curves.

\section{Methods}

\subsection{The curious case of the Sun}
The morphology of the solar light curve (as it would be observed in the Total Solar Irradiance or in the broad-band spectral passband like those of CoRoT, {\it Kepler}, or TESS) changes significantly depending on the phase of the solar activity cycle. While it appears to be quite regular at 11-year cycle minima when activity is low, the regularity disappears at periods of intermediate and high solar activity \citep{Lanza_and_Shkolnik2014, Aigrain2015, He2015}. In particular, \citet{Eliana2020} showed that if the Sun were observed by {\it Kepler}, the standard frequency analysis tools would most probably fail to detect the correct rotation period (unless observations are done during epochs of low solar activity). The causes for this inability are manifold: solar rotational variability is mainly brought about by spots (see, e.g., \citealt{Shapiro2016}). The relatively short lifetimes of sunspots from days to weeks (see, e.g., \citealt{Sami_spots}) implies that most of the spots transit the visible solar disk only once, which leads to irregularities in the solar light curve, and hampers the detection of the solar rotation period. Furthermore, the brightness changes of dark spots and bright faculae partly compensate each other, which decreases the amplitude of the rotational signal, further hindering the period determination \citep{Shapiro2017,Witzke2020,Nemec2020}. The exception from this general tendency are epochs of low solar activity with a small number of active regions. At these times the rotational variability is attributed to long-lived facular features and the light curve pattern becomes more periodic.

\subsection{The model}\label{model}
While the irregularity of the solar light curve is quite well understood, the situation gets more complicated for other early G-type stars. Their light curves look different, partly because the stars are observed at various inclinations. For example, faculae appear brighter at the limb and therefore contribute more strongly to the variability when the star is observed out of the ecliptic plane \citep{Nemec2020}. Additionally, stellar metallicity $\rm [Fe/H]$ affects facular (and to a smaller degree spot) contrasts \citep{Witzke2018}, which eventually has an impact on the period detectability \citep{Witzke2020}.

To synthesize the light curves of solar-like stars, we built on recent calculations by \cite{Nemec2020} and by \cite{Witzke2018}. \cite{Nemec2020} utilized a semi-empirical sunspot-group record by \cite{Jiang2011_1} and the Surface Flux Transport Model by \cite{Cameron2010} to reconstruct the distribution of active regions on the solar surface from the year 2010 back to 1700 with a daily cadence. By applying an appropriate geometrical transformation, \citet{Nemec2020} calculated the distribution of active regions on the solar disk as it would be observed at arbitrary inclinations. \cite{Witzke2018} calculated the brightness contrasts of faculae and spots relative to the quiet Sun (i.e. free from any apparent manifestations of magnetic activity) as a function of wavelength and position on the visible disk for stars with different metallicities and solar effective temperature. 

All in all, by combining the reconstructed disk distribution of active regions with their brightness contrasts, we generated light curves with a time span of 310~years as they would be seen in the passband of the {\it Kepler} telescope. The light curves were calculated for ten inclination angles $0^\circ \leq i \leq 90^\circ$ (with a step of $10^\circ$), and nine different metallicities $\rm -0.4 \leq [Fe/H] \leq 0.4$~dex (with a step of 0.1~dex). The solar record from 1700--2010 covers epochs of both low solar activity (like the Dalton minimum around 1790--1830), and very high solar activity (like the modern grand maximum around 1950--2000, see \citealt{Solanki2004, Usoskin2007}), which allows studying rotation period detectability during activity cycles of very different strengths. We note that, by assuming a solar disk distribution of active regions, we only account for the metallicity effect on the contrasts of magnetic features. A change in metallicity also affects the depth of the convective zone, which in turn could influence the stellar dynamo, in  particular the length of the stellar activity cycle or the emergence latitudes of magnetic bipoles \citep{Schuessler1992}. However, this effects is rather weak, e.g. doubling the metallicity of a star with solar temperature will deepen the convective zone by only ca. 8\% \citep{vanSaders2012,Karoff2018}.
Therefore, we expect these effects to be relatively small. Studying them is beyond the scope of the present paper, but would be an interesting future exercise.

\subsection{Monte Carlo approach}\label{MCMC}
We take a Monte Carlo approach to analyze light curves with different realizations of inclination angles and metallicities. The distribution of inclination angles is uniform in $\cos{i}$, where $i=0^\circ$ denotes a pole-on view and $i=90^\circ$ an equator-on view. The input distribution of metallicities was adapted for solar-like stars in the \textit{Kepler} field (see Fig.~\ref{input_dist} and \citealt{Reinhold2020}). For each (random) parameter combination $(i,[{\rm Fe/H}])$, we chose the model light curve from the grid with the closest parameters in metallicity and inclination angle. 

Following the observing strategy of the \textit{Kepler} telescope, we pick a random 4-year segment of the full time series (see top row of Fig.~\ref{lc_LPH_morph} for example). This light curve is then cut into 90-day segments (i.e. similar to the \textit{Kepler} observing quarters), where each 90-day chunk is normalized by its median, and appended to the previous one, to form a 4-year time series. These \textit{Keplerized} light curves (Fig.~\ref{lc_LPH_morph}, middle row) will be analyzed for rotation in the next step. The detrending is necessary because it filters out brightness variations on the activity-cycle timescale, and renders the light curves comparable to detrended {\it Kepler} data.

In addition to the various inclination and metallicity combinations, we study the impact of noise on the period detectability. The model light curves are by definition noise-free. In real observations, the visual stellar magnitude defines the noise level. We use the distribution of {\it Kepler} magnitudes $Kp$ of solar-like stars to compute different noise realizations $\sigma$ (see \citealt{Reinhold2020} for details). A noise time series with zero mean and standard deviation $\sigma$ is then added to the time series in the Monte-Carlo simulation. In total, we conducted 50,000 Monte Carlo runs, both for the noise-free and the noisy cases to study them separately.

\section{Results}

\subsection{Period detection}\label{period_detection}
\begin{figure}
  \centering
  \includegraphics[width=\textwidth]{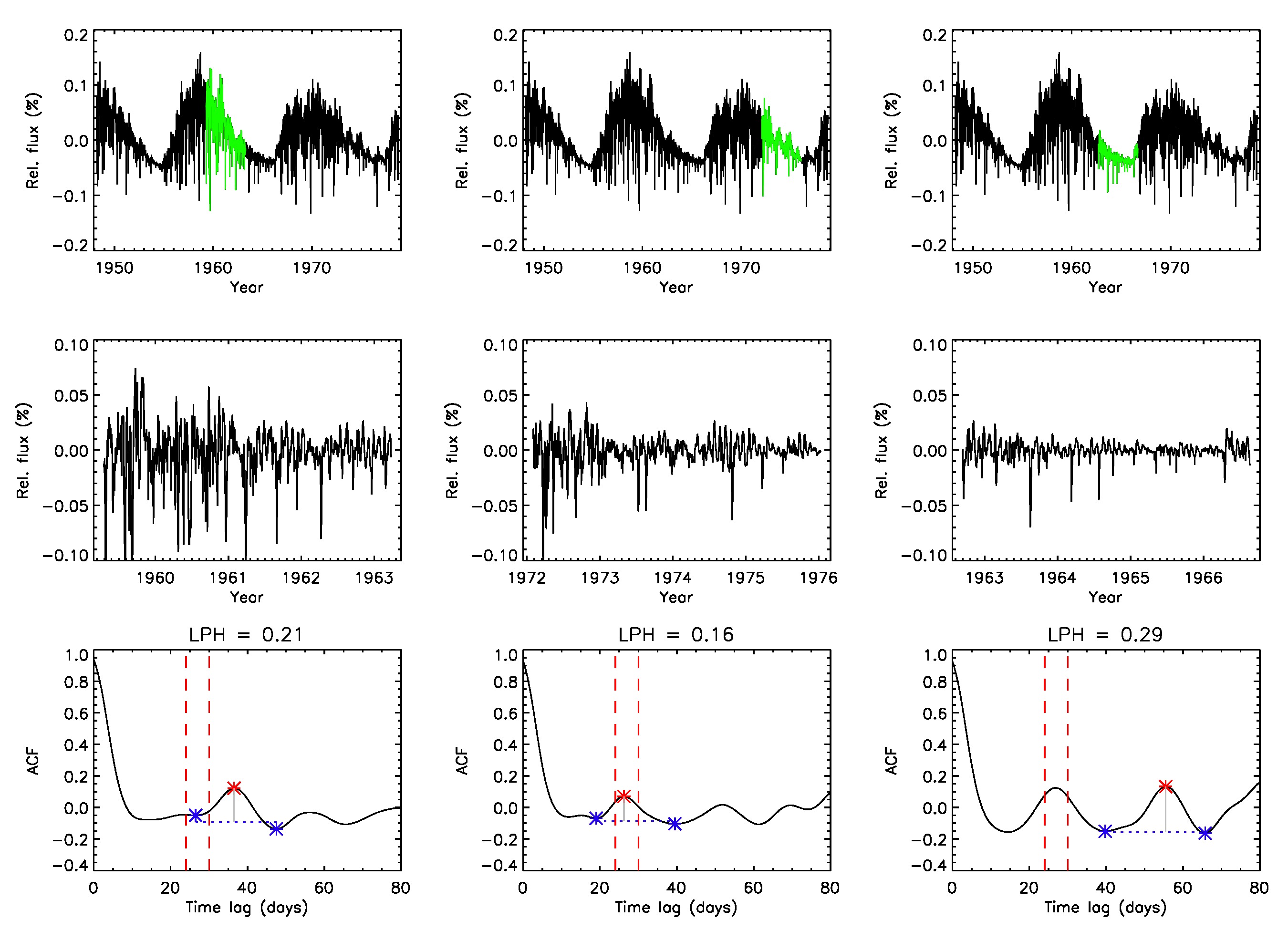}
  \caption{Top row: Model light curve (black) with an inclination of $i=40^\circ$ and solar metallicity $\rm [Fe/H]=0$, with three randomly chosen 4-year segments (green). Middle row: \textit{Keplerized} light curves of the chosen segment from the top row
  (see Sect.~\ref{MCMC} for details). Bottom row: auto-correlation function (ACF) of the selected 4-year segment. The measured period is indicated by the red asterisk, and the local peak height (LPH) is shown as the vertical gray line between the peak and the two troughs on either side. The vertical dashed red lines indicate the period detection window from 24 to 30~days.}
  \label{lc_LPH_morph}
\end{figure}

From among the various period detection methods, we chose the auto-correlation function (ACF) to search for periodicity in the time series (i.e. the same method as employed by \citetalias{McQuillan2014}). The ACF returns peaks of different power as a measure of the periodicity in the light curve. To quantify the strength of the periodicity, we adapt the measure of \citetalias{McQuillan2014}, where the local peak height (LPH) is computed as the difference between the highest peak and the mean of the two troughs on either side (see Fig.~\ref{lc_LPH_morph}, bottom row). We only search for peaks at periods less than 70~days, consistent with \citetalias{McQuillan2014}. If the highest ACF peak lies between 24--30~days and $LPH>0.1$, we count it as a detection. If the peak lies outside this period range or is smaller ($LPH<0.1$), it is counted as a false or non-detection.

Fig.~\ref{lc_LPH_morph}  illustrates the difficulty of detecting the correct rotation period of the Sun from the photometric time series obtained in the {\it Kepler} passband. The top row of Fig.~\ref{lc_LPH_morph}  shows the modeled light curve computed for a star with  solar metallicity, $\rm [Fe/H]=0$, as it would be observed outside of its equatorial plane at $i=40^\circ$. The bottom row gives the ACF and the computed LPH for three different 4-year segments of the same light curve. Depending on the selected segment, the ACF shows the highest peak at different periods. The first panel shows a peak with a moderate LPH but outside the range of 24--30~days (red dashed lines), i.e. a false detection. The second panel shows a peak close to the model rotation period of 27~days, although with a rather low LPH. The last panel shows a clear peak within the range 24--30~days, although this period is found to be the first harmonic of the highest peak at twice the correct rotation period (so that this panel corresponds to a false detection again). The light curve segment shown in this panel corresponds to an epoch of relatively low magnetic activity when the rotational variability of the Sun becomes faculae-dominated. Since faculae have significantly longer lifetimes than spots, this segment shows a more stable periodicity but even in such cases the correct rotation period is not necessarily associated with the highest ACF peak.

\begin{figure}
  \centering
  \includegraphics[width=\textwidth]{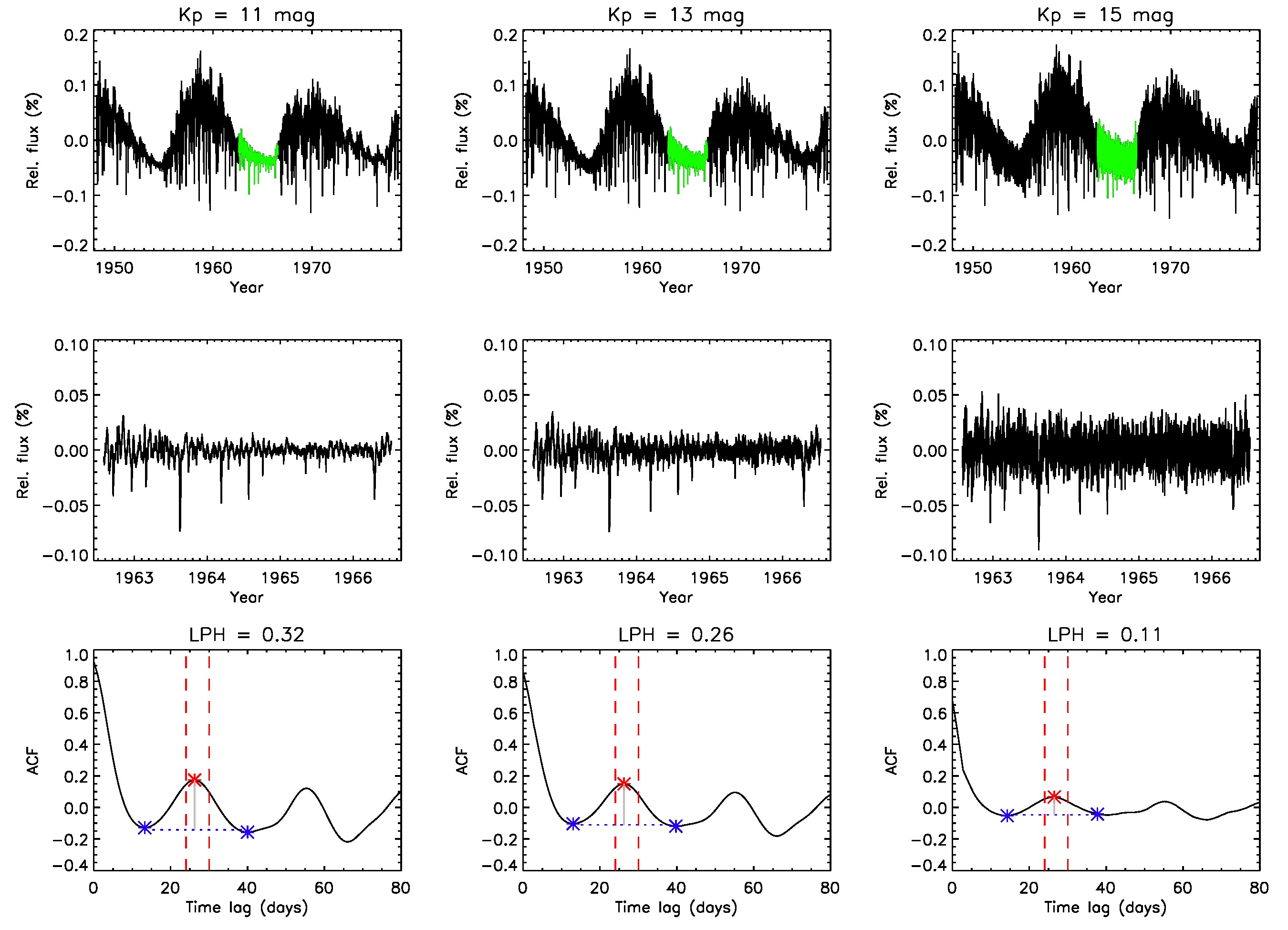}
  \caption{Same as Fig.~\ref{lc_LPH_morph} but choosing the same light curve segment for three different noise realizations corresponding to 11th (left), 13th (middle), and 15th (right) {\it Kepler} magnitude. The LPH decreases towards fainter stars.}
  \label{lc_LPH_noise}
\end{figure}

We now consider how the apparent magnitude of a star affects the period detection. For that purpose, Fig.~\ref{lc_LPH_noise} shows the same light curve as Fig.~\ref{lc_LPH_morph}, but with different noise levels to simulate the star as observed at different magnitudes. To demonstrate the effect of noise on the ACF, we chose a segment during solar minimum\footnote{The chosen segment slightly differs from the one chosen in the third panel Fig.~\ref{lc_LPH_morph} to demonstrate the effect of noise on the LPH.} where the correct period was detected, and added Poisson noise to the light curve, representative of a solar-like star at 11th, 13th, and 15th {\it Kepler} magnitude (see \citealt{Reinhold2020} for details). In all cases, the correct period was detected. While from 11th to 13th magnitude the LPH only slightly decreases, it decreases by more than half at 15th magnitude. 

\subsection{LPH dependence on spot area}
\begin{figure}
  \centering
  \includegraphics[width=\textwidth]{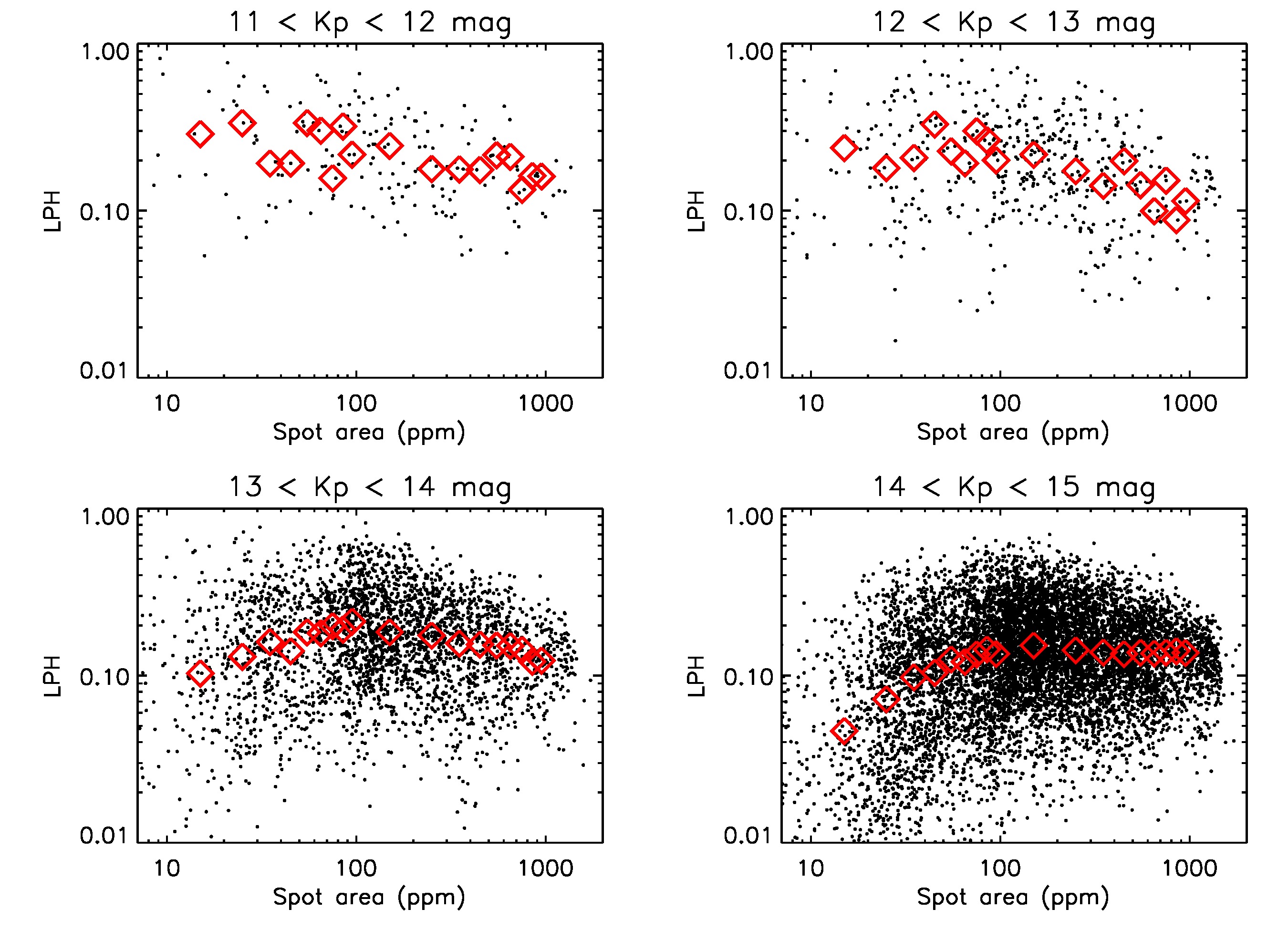}
  \caption{Local peak height (LPH) vs.~spot area fraction for different \textit{Kepler} magnitudes. The red diamonds show the median LPH values for the selected spot area bins. The very few peaks with $\rm LPH<0.01$ were excluded from the analysis.}
  \label{area_LPH_kepmag}
\end{figure}

The two examples in Sect.~\ref{period_detection}  illustrated how the period detection is affected by the activity level (Fig.~\ref{lc_LPH_morph}) and the amount of observational noise (Fig.~\ref{lc_LPH_noise}). Fig.~\ref{area_LPH_kepmag} combines both of these effects by showing the LPH when the highest peak was found between 24--30~days, as a function of the sunspot coverage on the visible solar disc, averaged over the 4-year segment (in ppm), for different magnitudes. The red diamonds show the median LPH values for the selected spot area bins to better illustrate the LPH dependence.

The upper left panel shows only stars brighter than 12th {\it Kepler} magnitude (i.e. with small noise levels). The LPH increases with decreasing spot area. As mentioned above, small spot coverages are typically found during activity minima when variability becomes faculae-dominated. Consequently, the light curves become more regular. A similar trend is found for stars between 12th and 13th magnitude but with larger scatter (upper right panel). Between 13th to 14th magnitude (lower left panel), the noise level becomes comparable to the variability amplitude during epochs of small spot coverages. As a result, the LPH drops for small coverages and the increase of the LPH with decreasing spot area can only be identified down to spot areas of 100\,ppm. For the faintest stars down to 15th magnitude (lower right panel), the larger noise further decreases the LPH for small spot areas, and for spot areas above 100\,ppm no trend can be identified any longer.

\subsection{Period distribution}
\begin{figure}
  \centering
  \includegraphics[width=\textwidth]{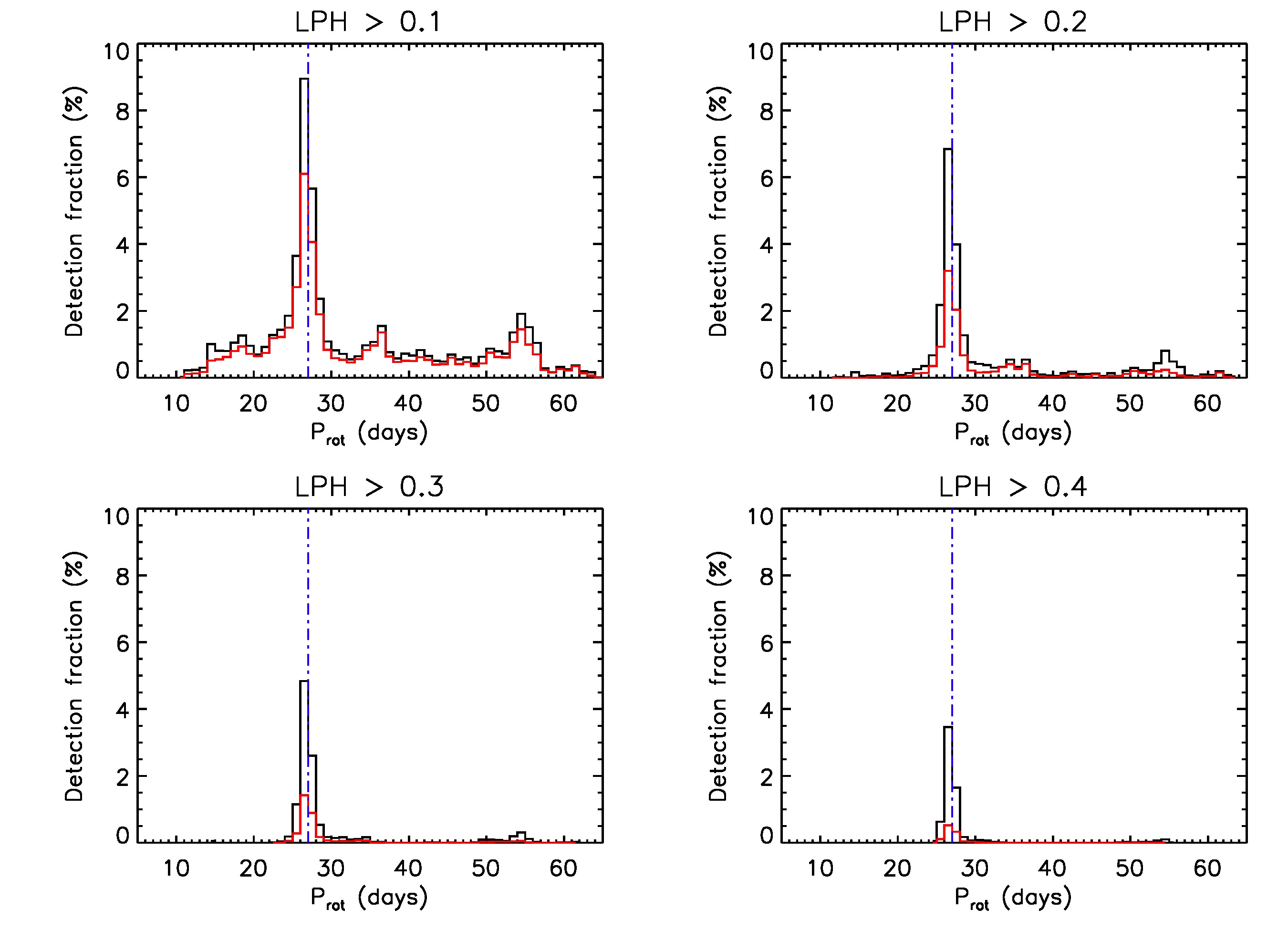}
  \caption{Rotation period distribution for different local peak heights of the noise-free (black) and noisy (red) cases. The blue dashed-dotted line indicates the model rotation period of 27~days.}
  \label{Prot_dist}
\end{figure}

As shown in the previous section, the position of the highest peak and the associated LPH determine the period detection. The percentages of correct and false detections are given in Table~\ref{det_table}. The period distribution is shown in Fig.~\ref{Prot_dist} for different LPH constraints for the noise-free (black) and the noisy (red) case. From the upper left to the lower right panel, the LPH threshold increases from 0.1 to 0.4. Consequently, the detection fraction decreases, but also the number of false detections drops. The decrease of detections is even stronger for the noisy case. We note that also the false detections decrease more strongly for the noisy case (see Table~\ref{det_table}). This is caused by the fact that a peak is more easily found in the noise-free case, but the associated period lies outside the range of 24--30~days, and is therefore counted as false detection.

When measuring rotation periods in real data, the period is a priori unknown, and one has to assign a certain LPH threshold, for which periods are considered as significant. The upper left panel in Fig.~\ref{Prot_dist} shows that even for small values ($0.1<LPH<0.2$), most detections are found at the correct (model) rotation period of 27~days. However, the number of false detections is also quite high (see Table~\ref{det_table}). Further increasing the LPH threshold significantly decreases the number of false detections but also lowers the number of correct detections. Finding an optimal LPH threshold that compromises between discarding correct detections and not having too many false periods is non-trivial. We stress that \citetalias{McQuillan2014} required $LPH>0.3$ to count the period as a real detection. As seen in the lower left panel, this threshold eliminates almost all false detections but strongly decreases the number of real detections (see discussion below).

\begin{table}
    \begin{tabular}{c|cc|cc}
\hline\hline
LPH & \multicolumn{2}{c|}{Detection} &  \multicolumn{2}{c}{False detection} \\
 & Noise-free & Noisy & Noise-free & Noisy \\
\hline
$>$0.1 & 23.8\% & 17.3\% & 35.2\% & 27.4\% \\
$>$0.2 & 15.5\% & 7.4\% & 9.0\% & 4.5\% \\
$>$0.3 & 9.5\% & 2.9\% & 1.9\% & 0.5\% \\
$>$0.4 & 6.1\% & 1.0\% & 0.4\% & 0.0\% \\
\hline
\end{tabular}

    \caption{Detections and false detections for different LPH for both the noise-free and noisy cases.}
    \label{det_table}
\end{table}

\subsection{Detection rate}
\begin{figure}
  \centering
  \includegraphics[width=0.5\textwidth]{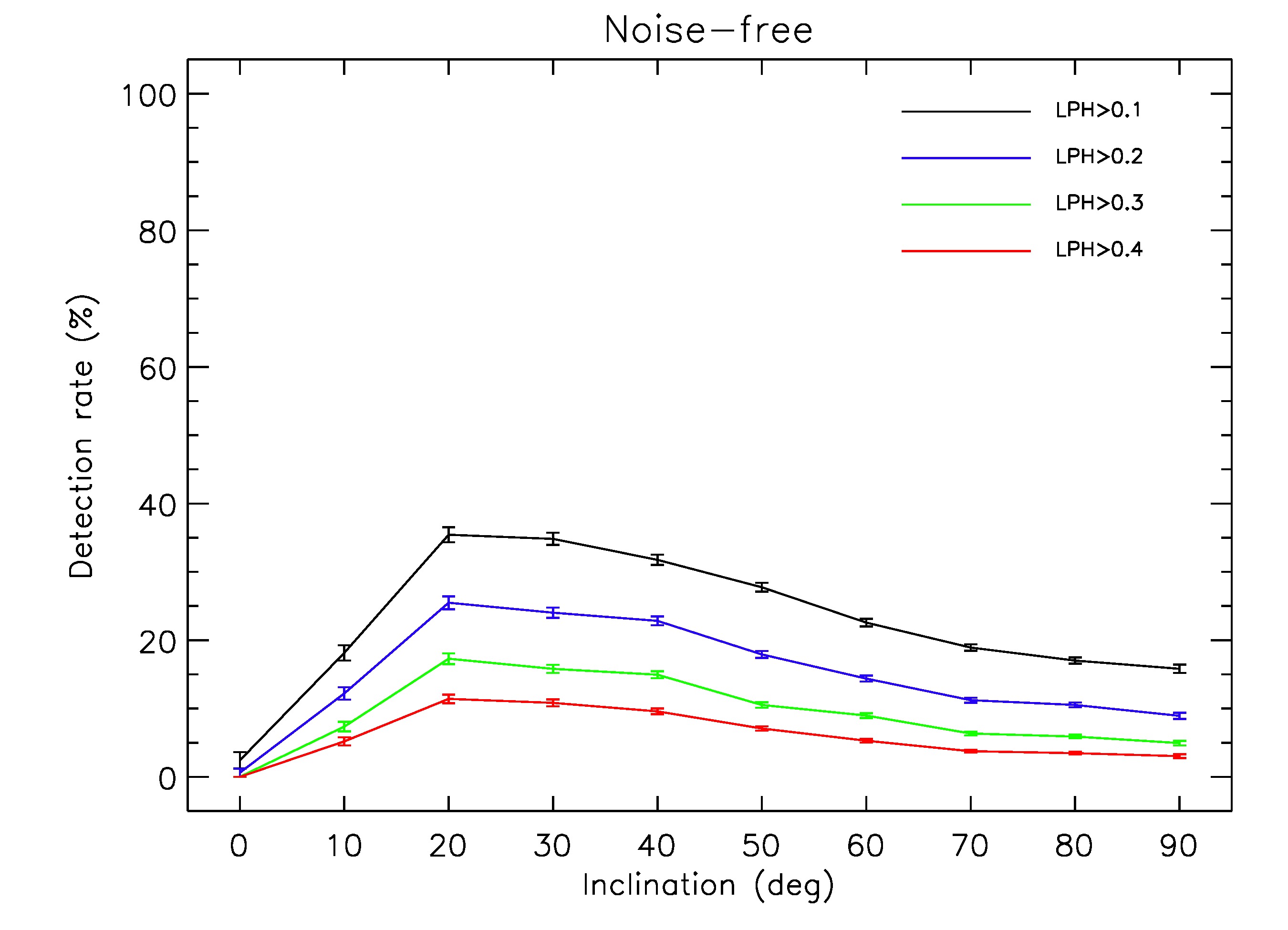}\includegraphics[width=0.5\textwidth]{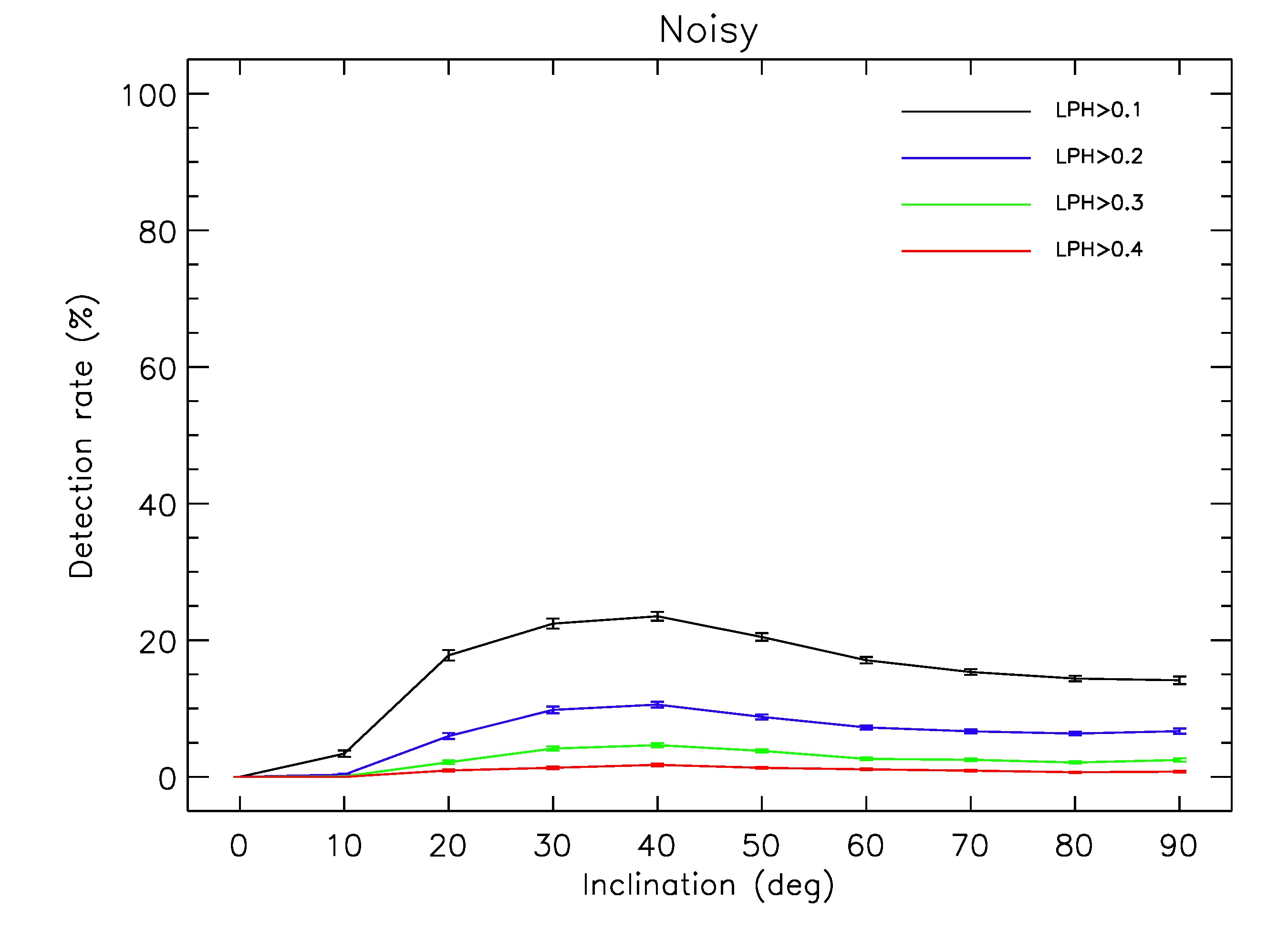}
  \caption{The detection rate as a function of inclination angle, integrated over all metallicities, for different LPH thresholds. The error bars show the square root of the number of detections divided by the number of models at a given inclination.}
  \label{det_rate_inc}
\end{figure}

We now turn to the question how the period detectability is affected by stellar inclination and metallicity. For that purpose, we define the detection rate as the number of detections divided by the number of different Monte Carlo runs at a given parameter. In Fig.~\ref{det_rate_inc} we show the detection rate as a function of the inclination of the rotation axis of the model star (integrated over all metallicities) for different LPH values. The error bars indicate the square root of the number of detections divided by the number of models. As before, we consider the noise-free (left panel) and the noisy (right panel) cases separately.

The noise-free case qualitatively displays the same behavior for all LPH thresholds. As expected, the detection rate is zero for the pole-on view. However, when increasing the inclination angle the detection rate steeply increases to the maximum at an inclination near $20^\circ$, and gradually decreases towards the equator-on view at $90^\circ$. This result might be surprising at first glance but can be explained by the dominant contribution of faculae to brightness variability for stars with near-equatorial activity belts (similar to those on the Sun) observed close to the pole-on view \citep{Shapiro2016}. In such stars each facular feature spends roughly half a rotation period on the far-side of a star and the remaining half of the time near the limb on the visible disk. Faculae appear especially bright near the limb, and usually last for several stellar rotations. Consequently, the light curves of such stars appear more regular, leading to higher LPH values. We note that the calculations are performed assuming a solar latitudinal distribution of active regions. A change of the distribution would affect the visibility of the active regions, and consequently, the inclination angle corresponding to the maximum of the detection rate. However, we expect that a solar distribution is typical for stars with solar rotation period and temperature (see Sect.~\ref{model}).

In the noisy case (right panel) the detection rates are generally smaller (cf. Table~\ref{det_table} and Fig.~\ref{Prot_dist}). While the curves have a similar shape to the noise-free cases, their maxima are shifted to higher inclinations. Such a shift is caused by the decrease of the amplitude of brightness variability with decreasing inclination \citep{Nemec2020}, and consequently a decrease of the signal-to-noise ratio. Consequently, the detection peak near $20^\circ$ is suppressed, leaving a residual peak near $40^\circ$. As already shown in Figs.~\ref{area_LPH_kepmag} and \ref{Prot_dist}, the noise decreases the LPH such that only a few cases with $LPH>0.3$ remain. 

\begin{figure}
  \centering
  \includegraphics[width=0.5\textwidth]{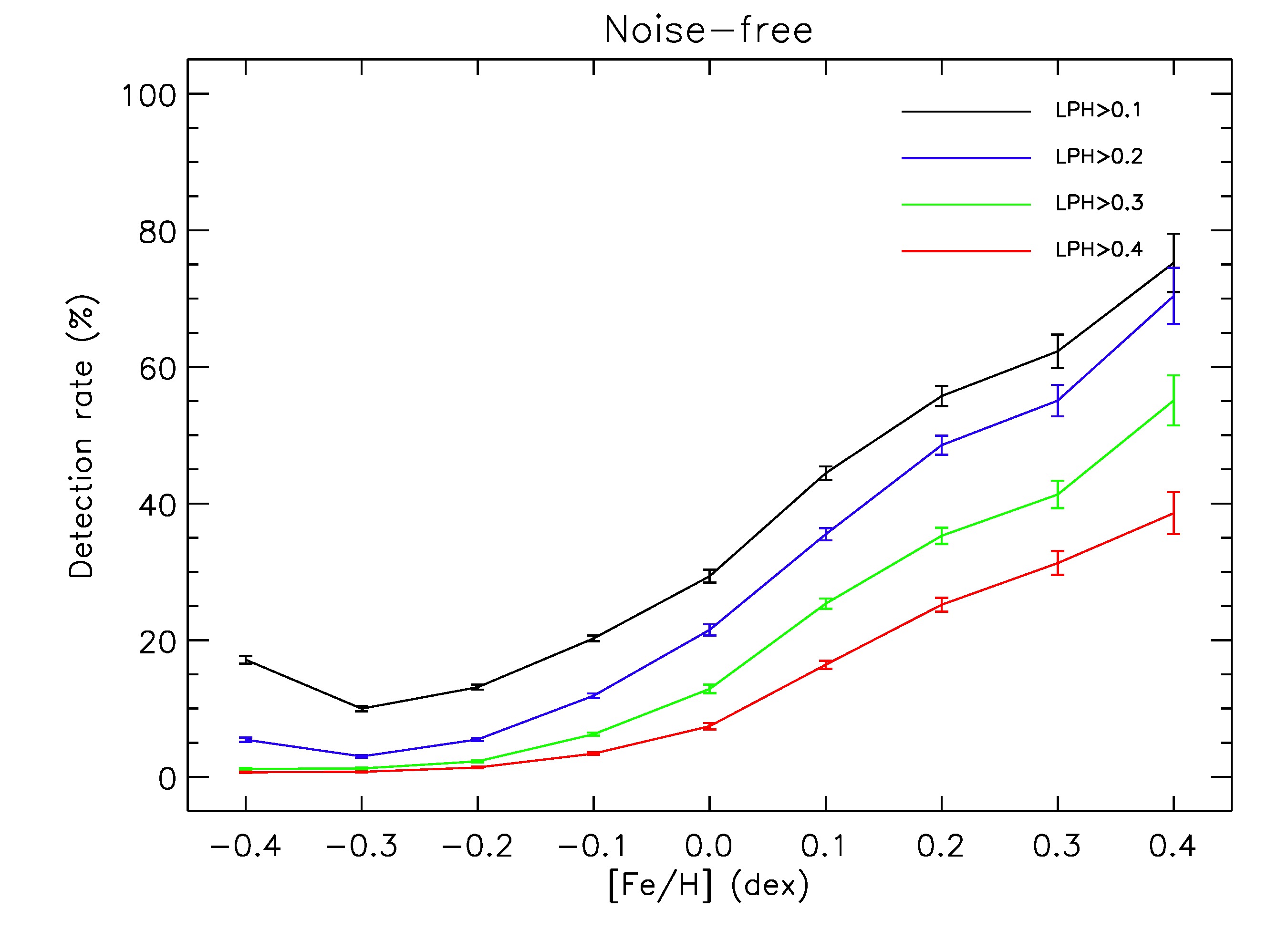}\includegraphics[width=0.5\textwidth]{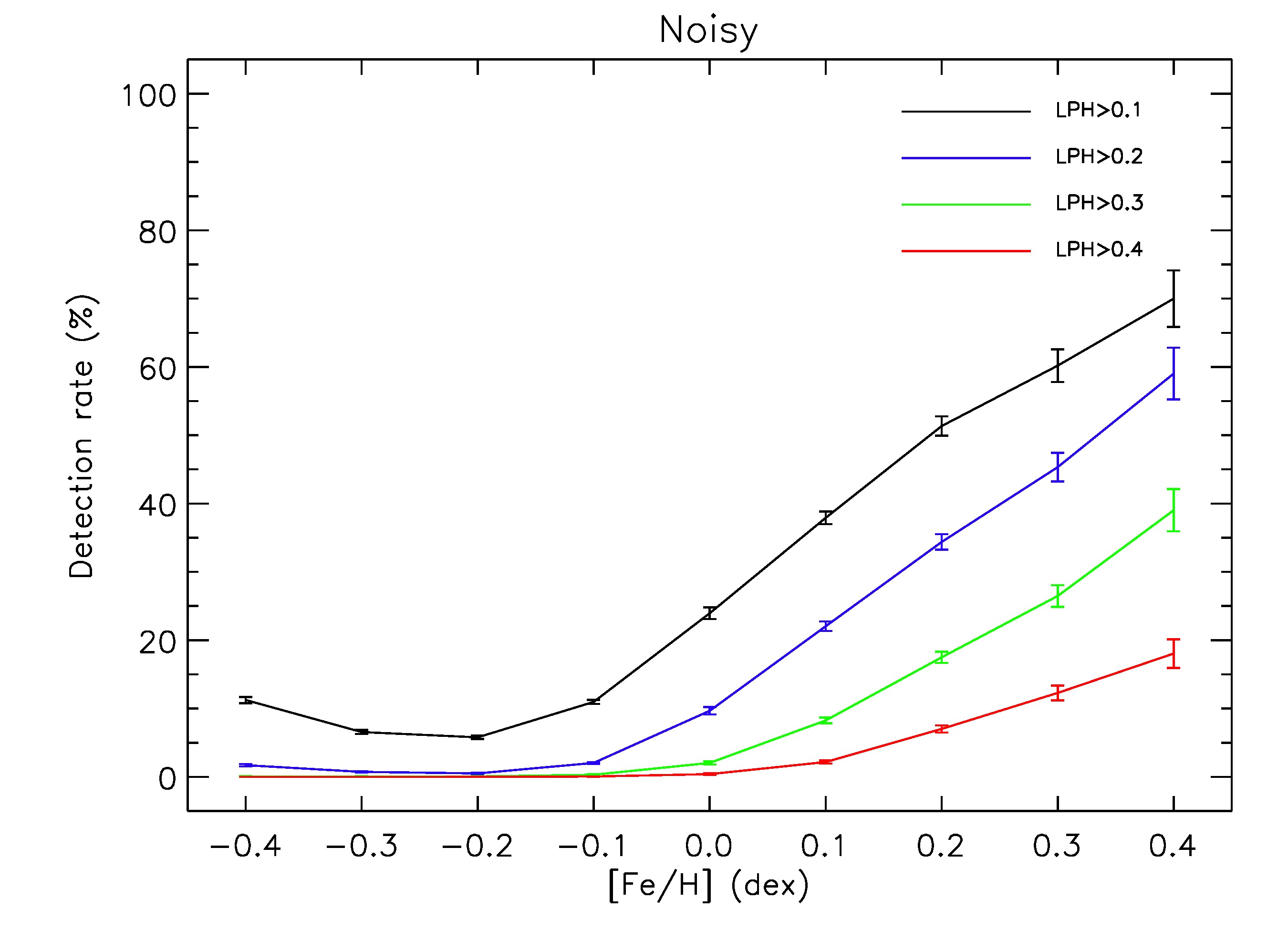}
  \caption{The detection rate as a function of metallicity, integrated over all inclinations, for different LPH thresholds. The error bars show the square root of the number of detections divided by the number of models at a given metallicity.}
  \label{det_rate_FeH}
\end{figure}

In Fig.~\ref{det_rate_FeH} we show the detection rate as a function of metallicity (integrated over all inclinations) for different LPH values. We note that the qualitative shape of the $LPH>0.1$ curve is consistent with the one found in \citet{Witzke2020}, who considered the noise-free case (see Sect.~\ref{intensities} in the appendix for the difference between the calculations in this study and those employed in \citealt{Witzke2020}). Fig.~\ref{det_rate_FeH} shows that for both the noise-free (left panel) and the noisy (right panel) case, the detection rate increases with metallicity. This is caused by the stronger contribution of the faculae to the overall variability. Only for the cases $LPH>0.1$ and $LPH>0.2$, does the detection rate show a minimum at $\rm [Fe/H]=-0.3$~dex or $-0.2$~dex (noisy), increasing again towards smaller metallicity values. We expect that this trend continues towards even smaller metallicities. 

\subsection{Comparison with observations}
We now compare our detection rates (see Table~\ref{det_table}) to period detections of solar-like stars in the \citetalias{McQuillan2014} sample. As mentioned above, \citetalias{McQuillan2014} used a relatively conservative detection threshold of $LPH>0.3$. The bottom left panel of Fig.~\ref{Prot_dist} and Table~\ref{det_table} indicate that this threshold represents only the tip of the iceberg: for $LPH>0.3$ (noisy case), the rotation periods can be correctly detected for only 2.9\% of our modeled light curves.

Galactic evolution models \citep{vanSaders2019} predict that 16\% of the (dwarf) stars in the {\it Kepler} field with effective temperatures $5500 < T_{\rm eff} < 6000$\,K should have rotation periods between 24--30~days (van Saders, private communications). Using the latest {\it Kepler} parameter catalog \citep{Mathur2017}, we select stars in this temperature range, with surface gravities $\log g>4.2$ to exclude more evolved stars, and brighter than 15th {\it Kepler} magnitude (following the selection criteria used in \citealt{Reinhold2020}). We further restrict the catalog metallicities to $\rm -0.45<[Fe/H]<0.45$~dex, which corresponds to the range of simulated metallicities (see Fig.~\ref{input_dist}). Selecting such stars from tables~1 and 2 in \citetalias{McQuillan2014} yields $N=16890$ stars. Among those, only 16\% will have periods between 24--30~days, and according to our analysis, only 2.9\% of these stars will have \textit{detectable} periods. Thus, we estimate that $N_{det}=N*0.16*0.029=78$ stars should have detectable periods.

\citetalias{McQuillan2014} found 455 stars in this parameter range with periods $24 \leq P_{\rm rot} \leq 30$~days. However, the vast majority of these stars exhibits variability levels much higher than the Sun, and represent a regime of variability very different from that of the Sun \citep{Reinhold2020, Jinghua2020, Emre2020}. Consequently, the light curves of these stars cannot be accounted for by our model. To correct for such stars, we followed the approach of \cite{Witzke2020} and selected the stars with variability (regressed to solar values of effective temperature, metallicity, and rotation period, see Fig.~S8 and its detailed discussion in \citealt{Reinhold2020}) below 0.18\%, which corresponds to the maximum variability of the Sun over the last 140~years. All in all, only 73 out of the 455 stars satisfied such a criterion. This number is gratifyingly close to our estimate of 78 stars.

\section{Conclusions}
In this study we identified biases in the period determination of stars with solar-like variability. The detection rates among these stars are lower than for stars of other spectral types. In particular, only 2.9\% of them would be detectable using the thresholds set in \citetalias{McQuillan2014}. This is mainly caused by the small variability amplitudes of the rotational tracers and their relatively short lifetimes compared to the rotation period. 

The very low detection rate explains the large discrepancy between the number of measured rotation periods \citepalias{McQuillan2014}, and those predicted by Galactic evolution models \citep{vanSaders2019}. The predicted number of stars with detectable periods (78), and that for which rotation periods have actually been measured (73), is remarkably similar. Fig.~\ref{Prot_dist} shows that many more rotation periods of solar-like stars may be measured when lowering the thresholds in the automated period surveys. However, this will also add a number of false periods, depending on how the thresholds are set. 

Our study revealed that the rotation periods of most solar-like stars will go undetected using standard frequency analysis tools. Thus, we emphasize the importance of alternative methods for period detection such as the GPS method \citep{GPS_I,Eliana2020} or new approaches based on Gaussian process regression \citep{Foreman-Mackey2017,Angus2018,Kosiarek2020}.

\acknowledgements
We would like to thank Jennifer van Saders for providing model numbers and for helpful discussion. 
This work has received funding from the European Research Council (ERC) under the European Union's Horizon 2020 research and innovation programme (grant agreement No. 715947). This work has been partially supported by the BK21 plus programme through the National Research Foundation (NRF) funded by the Ministry of Education of Korea.  We would like to thank the International Space Science Institute, Bern, for their support of science team 446 and the resulting helpful discussions.

\appendix
\counterwithin{figure}{section}

\section{Monte Carlo input distribution}
\begin{figure}
  \centering
  \includegraphics[width=\textwidth]{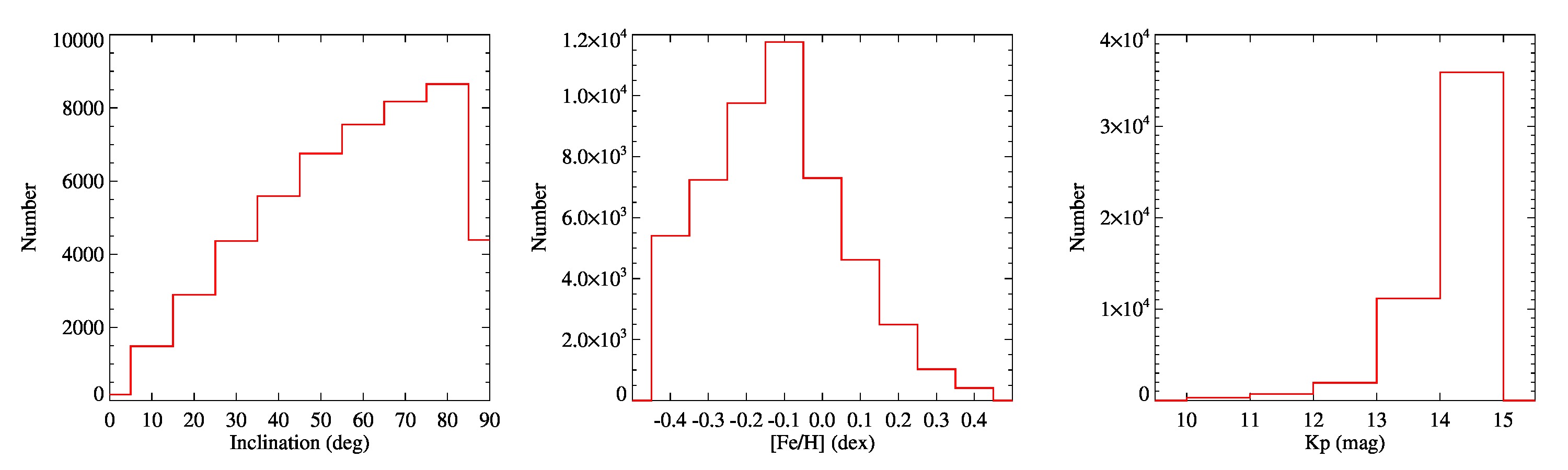}
  \caption{Input distributions of inclination angles (left) and metallicities (middle), and {\it Kepler} magnitudes (right) of the Monte Carlo simulation.}
  \label{input_dist}
\end{figure}

Fig.~\ref{input_dist} shows the distributions of input parameters used in the Monte Carlo simulation. The first panel shows the distribution of inclination angles. It can be shown that isotropic inclination angles $i$ exhibit a uniform distribution in $\cos i$ (see e.g. \url{http://keatonb.github.io/archivers/uniforminclination} for a detailed derivation). The last bin of the distribution ($85-90^\circ$) only contains half the number of realizations because no inclination angles greater than $90^\circ$ exist. The same argument applies to the first bin from $0-5^\circ$.

The middle panel shows the distribution of metallicities of the solar-like stars in the {\it Kepler} field. The catalog values were adapted from \citet{Mathur2017} and the selection of solar-like stars can be found in \citet{Reinhold2020}. We note that the Sun ($\rm [Fe/H]=0$) is slightly more metal-rich than the peak of the distribution.

The last panel shows the apparent magnitudes of the stars in the {\it Kepler} field. It is obvious that the majority of stars is very faint. Since the stellar magnitudes define the noise in the light curves, it is crucial to adapt this distribution for the noise model (see \citealt{Reinhold2020}) to make realistic predictions about stars in the {\it Kepler} field.

\section{Generating light curves}
The total spectral flux at a certain time is composed of fluxes emerging from surface areas with different levels of magnetic activity. Following the detailed description in \citet{Shapiro2014A&A}, we decompose the spectral flux from a star, $ F$, into contributions from the quiet stellar region ($ F_{\rm Q}$), faculae ($F_{\rm fac}$), and spots ($F_{\rm spot}$):   
\begin{equation}
\label{eq:total_flux}
F(\lambda) =  F_{\rm Q}(\lambda) + F_{\rm fac}(\lambda) + F_{\rm spot}(\lambda),
\end{equation}
where $\lambda$ is the wavelength. For the quiet stellar region, the disc-integrated flux $\rm F_Q(\lambda)$ is obtained by integrating  
\begin{equation}
F_Q (\lambda) = \int_0^1 I_Q(\lambda, \mu) \omega(\mu) d\mu,
\end{equation}
where $\omega(\mu) = 2\pi \mu (r_{star}/ d_{star})^2$ is a weighting function with the stellar radius, $r_{star}$, and the distance between the star and the observer, $d_{star}$. The emergent intensity, $I_Q(\lambda, \mu )$,  also depends  on $\mu$, which is the cosine of the angle between the observer's direction and the local stellar radius. In this formulation the stellar disc center is associated with $\mu =1$ and the limb with $\mu =0$.

Both faculae and spots are taken into account through their contrast with respect to the quiet regions. Therefore, the contribution of faculae is defined as
\begin{equation}
\label{eq:Faculae_flux}
F_{Fac}(\lambda )= \int_0^1  \alpha_{Fac}(\mu) \left[I_{Fac}(\lambda, \mu) - I_Q (\lambda, \mu) \right]\omega(\mu) d\mu,
\end{equation} 
where the fractional coverage of the ring corresponding to a given $\mu$ by faculae is given by the function $\alpha_{Fac}(\mu)$.

The contribution  from spots consists of those from spot umbrae and spot penumbrae:
\begin{eqnarray}
\label{eq:Spot_flux}
F_{\rm spot}(\lambda )&=&    \int_0^1 \alpha_{\rm pen}(\mu) \left(I_{\rm pen}(\lambda, \mu) - I_Q (\lambda, \mu) \right)  \omega(\mu) d\mu   \nonumber \\
 &+ & \int_0^1    \alpha_{\rm umb}(\mu) \left(I_{\rm umb}(\lambda, \mu) - I_Q (\lambda, \mu) \right)   \omega(\mu) d\mu, 
\end{eqnarray}
where $I_{\rm umb}$ and $I_{\rm pen}$ are the emergent intensities  from the spot umbrae and spot penumbrae, respectively, and the $\alpha_{\rm umb}$ and $\alpha_{\rm pen}$ denote the corresponding surface coverages. 

The surface coverages for the magnetic features (i.e. $\alpha_{\rm Fac}$, $\alpha_{\rm umb}$  and $\alpha_{\rm pen}$) used in this work are taking from \citet{Nemec2020}. Furthermore, calculations of the emergent intensities for all stellar regions follow the approach used in 
\citet{Witzke2020}, but with a small modification which is explained in Appendix~\ref{intensities}.

\section{Calculating emergent intensities}\label{intensities}
The model atmospheres and corresponding emergent spectra computed by  \citet{Unruh_Yvonne_1999} for the solar faculae, spots, and quiet regions (hereafter, original models) proved to be very successful in reproducing the solar brightness variations with high accuracy \citep{Krivova_2003A&AL, MPS_AA, TOSCA2013}.

Here, we extend the intensities for different surface components to different metallicities following the approach outlined in \citet{Witzke2018, Witzke2020}, but with slight modifications to cover a broader metallicity range. In our modeling approach, we aim to match the intensity contrasts for the solar metallicity as closely as possible to the original models. Thus, in the first step we searched for the model parameters (input parameters for calculating stellar atmospheres with ATLAS9), such as convection settings, surface gravity and continuum opacity sources for the quiet Sun, spot-umbra and spot-penumbra, to match the original models and spectra by \cite{Unruh_Yvonne_1999}. The closest match is achieved by the parameters listed in Table~\ref{tbl:model_params}. Then for the generation of the faculae model we assumed that the temperature difference, $\Delta \text{T}$, and  pressure difference, $\Delta P$, as a function of column mass between the original facular and quiet Sun models are the same as between our new facular and quiet Sun models.

Finally, to calculate atmospheric models for different metallicity values, we first generated atmosphere models for the quiet regions and the spots assuming radiative equilibrium. Then we followed up on the \citet{Witzke2018} approach and assumed that a change of the  metallicity value has the same effect on the temperature and pressure structures of the quiet Sun and faculae. Using the quiet stellar atmosphere models for different metallicities, we applied the solar $\Delta \text{T}$ and $\Delta P$ with column mass to calculate the facular models. Using these atmospheric models for the quiet regions and magnetic features, we generated the emergent intensities $I_{\lambda, \mu}$ for each metallicity value using the MPS-ATLAS code \citep{Witzke_et_al_2021}.

\begin{table}
    \begin{tabular}{c|c|c|c|c}
\hline\hline
model & effective &  surface  & mixing & over \\
      & temperature [K] & gravity &  -length & -shoot \\[5pt]
\hline
Quiet region  & 5777  & 4.43777    &  1.25     & on \\
Spot umbra    & 4500  &    4.0     &  1.25     & on  \\
Spot penumbra & 5450  &    4.0     &  1.25     & on  \\
\hline
\end{tabular}
    \caption{Input parameters for model atmospheres in radiative equilibrium.}
    \label{tbl:model_params}
\end{table}

\bibliography{references}
\bibliographystyle{aasjournal}

\end{document}